\documentclass[9pt,technote,twocolumn]{IEEEtran}

\IEEEoverridecommandlockouts                              % This command is only
\usepackage{multirow}
\usepackage{graphics,graphicx,epsfig,epic,eepic,amssymb,latexsym,amsmath,amssymb,amsmath,picinpar}
\usepackage{subfigure,color}
\usepackage{epstopdf}
\usepackage{url}

\newcommand{\Real}{\mathop{{\rm I}\mskip-4.0mu{\rm R}}\nolimits}

\newcommand{\beq}{\begin{equation}}
\newcommand{\eeq}{\end{equation}}
\newcommand{\beqar}{\begin{eqnarray}}
\newcommand{\eeqar}{\end{eqnarray}}
\newcommand{\beqarno}{\begin{eqnarray*}}
\newcommand{\eeqarno}{\end{eqnarray*}}
\newcommand{\ba}[1]{\begin{array}{#1}}
\newcommand{\ea}{\end{array}}

\newtheorem{itlemma}{Lemma}[section] %number by section (set in \em by default)
\newtheorem{itcorollary}[itlemma]{Corollary}
\newtheorem{itremark}{Remark}
\newtheorem{itdefinition}[itlemma]{Definition}
\newtheorem{itexample}[itlemma]{Example}
\newenvironment{remark}{\begin{itremark}\rm}{\end{itremark}} %no-italics

\newtheorem{problem}{{\bf{Problem}}}
\newtheorem{teorema}{{\bf{Theorem}}}
\newtheorem{proposition}{{\bf{Proposition}}}

\title{\LARGE \bf
Explicit Reference Governor for Continuous Time Nonlinear Systems Subject to Convex Constraints
\author{E. Garone, M. Nicotra % <-this % stops a space
\thanks{E. Garone and M. Nicotra are with the department SAAS of Universit\'{e} Libre de Bruxelles (ULB), 50 Av. F.D. Roosvelt, B-1050 Brussels, Belgium
    {\tt\small \{egarone, mnicotra\}@ulb.ac.be}}
}
}
\begin{document}
\maketitle
\thispagestyle{empty}
\pagestyle{empty}

%%%%%%%%%%%%%%%%%%%%%%%%%%%%%%%%%%%%%%%%%%%%%%%%%%%%%%%%%%%%%%%%%%%%%%%%%%%%%%%%
\begin{abstract}
\noindent
This paper introduces a novel closed-form strategy that dynamically modifies the reference of a pre-compensated nonlinear system to ensure the satisfaction of a set of convex constraints. The main idea consists of translating constraints in the state space into constraints on the Lyapunov function and then modulating the reference velocity so as to limit the value of the Lyapunov function. The theory is introduced for general nonlinear systems subject to convex constraints. In the case of polyhedric constraints, an explicit solution is provided for the large and highly relevant class of nonlinear systems whose Lyapunov function is lower-bounded by a quadratic form. In view of improving performances, further specializations are provided for the relevant cases of linear systems and robotic manipulators.
\end{abstract}
\vspace{-0.15 cm}
\section{Introduction}
\vspace{-0.15 cm}
A fundamental aspect that arises when controlling real plants is that the system must not only be stabilized asymptotically, but must also satisfy a set of constraints at all times. Several schemes have been proposed in the literature to deal with this issue, these approaches can be roughly grouped in two main families.

The first family consists of Model Predictive Control (MPC) \cite{Morari1999,MPC1,MPC2,MPC3,MPC4} schemes. From an industrial viewpoint, the adoption of this kind of solution requires to discard existing control laws and to close the loop using a control law based on online optimization.

An alternative solution is to provide an already existing controller, tuned for high performance in close proximity of the reference, with constraint handling capabilities for larger transients.  This second choice, although less performing than MPC solutions, may be attractive for practitioners willing to preserve existing controllers and/or to limit issues related with computational effort, tuning complexity, stability and robustness certification requirements.  Anti-windup schemes are classical ways to do so. Another way is the use of reference/command governors.

A reference governor is an {\em add-on}  scheme which enforces state and control constraints by acting on the reference of an existing closed-loop system. Several reference governors have been proposed in the literature. For a comprehensive discussion on these schemes please refer to the tutorial survey \cite{SurveyACC}.

Reference governors for linear systems were first proposed as continuous-time algorithms in \cite{Kapasouris90}. Later on, reference governors for the discrete-time framework \cite{Gilbert94,GilbertTan1991} have emerged due to some implementation advantages. Formulations of reference and command governors have appeared in \cite{Bemporad1994,Casavola2000,comgov,CaGaTe2013}.

For what concerns nonlinear systems, approaches using linearized models have been investigated and used in several real applications, see e.g. \cite{Kalabic2011b},\cite{Casavola2004}. Reference governors explicitly designed for nonlinear discrete time systems have also been proposed, see e.g. \cite{Angeli1999, Bemporad1998,Gilbert199a,Miller2000}. In \cite{Angeli1999b} some comparisons of direct nonlinear versus linearization-based schemes are reported.

As pointed out in \cite{nonlingov}, the common feature of the nonlinear schemes is that the reference is chosen (implicitly or explicitly) so that it belongs to a state-dependent admissible set. This admissible set must be built so as to ensure \emph{safety } (i.e. if an admissible reference is held constant, constraints are not violated) and \emph{strong returnability} (i.e. if an admissible reference is held constant, then after a fixed time the state dependent admissible set will contain again the reference). A way to implicitly build this kind of set is to make use online predictive simulations (see e.g. \cite{Bemporad1998}). Another approach, introduced in \cite{Gilbert199a} and \cite{refgov2}, exploits the use of Lyapunov functions. The main idea of these schemes is to convert state constraints into constraints on the value of the Lyapunov function. Then, making use of online optimization, the reference is selected so as to ensure that the one-step state prediction satisfies the bound on the Lyapunov function.

In this paper, we propose a reference governor for continuous time nonlinear system based on the idea of constraining the Lyapunov function. The main difference w.r.t. previous approaches is that, rather than manipulating directly the reference, the methods acts on its derivative. This allows to change the reference using an explicit control law and therefore avoid the use of explicit predictions and optimization procedures.
The properties of this method will be investigated together with a discussion on the computational aspects. It will be remarked that, in the general case, the main difficulty to overcome is the determination of the bounds for the Lyapunov functions. A general closed form solution will be provided for the highly relevant case of systems subject to polyhedral constraints and whose Lyapunov functions can be bounded by a quadratic form. Further specializations to the relevant sub-cases of linear systems and robotic manipulators will be provided, as well.
\section{System Description and Problem Statement \label{sec: ProblemStatement}}

Let
\begin{equation}\label{system}
\dot{x}=f\left(x(t),g(t)\right)
\end{equation}
describe the closed-loop dynamics of a pre-compensated system, subject to convex constraints
\begin{equation}\label{constraints}
c_{i}\left(x(t),x_{g(t)}\right)\geq 0,\; i=1,\ldots,n_{c}, \forall t.
\end{equation}
It is assumed that the system has been compensated so that $f(x,g)$ is Lipschitz and, for any constantly applied reference $g \in \Real^m$, the associated steady state $x_{g} \in \Real^n$ is Globally Asymptotically Stable (GAS). It is assumed that $x_{g}$ is a class $C^1$ function of $g$. \\

The Reference Governor (RG) problem can be defined as follows:

\begin{problem}\label{PROBLEM}
Consider the pre-compensated system (\ref{system}) subject to constraints (\ref{constraints}) and denote with $r(t)$ the desired reference signal (not known in advance). The RG design problem is that of generating, at each instant $t$, a suitable reference $g(t)$ such that:
\begin{itemize}
\item its application never leads to constraint violation, i.e., $c_{i}\left(x(t),x_{g(t)}\right)\geq 0,\; i=1,\ldots,n_{c}, \forall t \geq 0$
\item $g(t)$ approximates $r(t)$ as much as possible.
\end{itemize}
\end{problem}

\section{A General Lyapunov-based Reference Governor \label{sec: ProblemStatement}}
The strategy proposed in this paper is based on the observation that the Lyapunov function defines an invariant level set centered on the steady state $x_g$. To guarantee that the constraints are satisfied, it is therefore sufficient to manipulate the applied reference in such a way that the Lyapunov function remains smaller than a suitable upper bound. This idea has been introduced for the first time in \cite{Gilbert199a},\cite{refgov2}, where constraints were enforced for discrete time systems by solving at each time step a convex optimization problem. In this paper, a closed-form approach for continuous time systems is presented that acts on the first derivative of the reference and does not require any online optimization.
\\

Let $V\left(x,x_{g}\right)$ be a Lyapunov function such that its time derivatives is
\begin{equation}
\dot{V}\left(x,x_{g},\dot{g}\right) \leq 0\,\,\,\,\,\mathrm{for}\,\,\dot{g}=0, \forall x \in \Real^n,g\in \Real^m .
\end{equation}
$\dot{V}$ can be negative semidefinite if the LaSalle theorem can be used to prove GAS, otherwise it is required that $\dot{V}<0$.
\\
Moreover, assume that it is possible to compute a set of smooth functions $\Gamma_{i}\left(x_{g}\right)$ such that
\begin{equation}\label{conditionOnGammai}
V\left(x,x_{g}\right)\leq \Gamma_{i}\left(x_{g}\right)\quad\Rightarrow\quad c_{i}\left(x,x_{g}\right)\geq 0,
\end{equation}
meaning that $V\left(x,x_{g}\right)\leq \Gamma_{i}(x_{g})$ defines an invariant level-set centered in $x_g$ that is completely contained in the $i$-th constraint (see Figure 1 for a graphical depiction).
\\
By defining the set $\mathcal{I}$ as
\begin{equation}\label{Iset}
\mathcal{I}=\left\{ i:\;\Gamma_{i}\left(x_{g}\right)=\min_j\left(\Gamma_{j}\left(x_{g}\right)\right)\right\} ,
\end{equation}
all constraints (\ref{constraints}) can be verified simultaneously by enforcing at each time instant
\begin{equation}\label{LyapConstraints}
V\left(x(t),x_{g(t)}\right)\leq\Gamma_{\mathcal{I}}\left(x_{g(t)}\right).
\end{equation}

\begin{figure}
\label{Gammaiii}
\centering
\includegraphics[width=8cm]{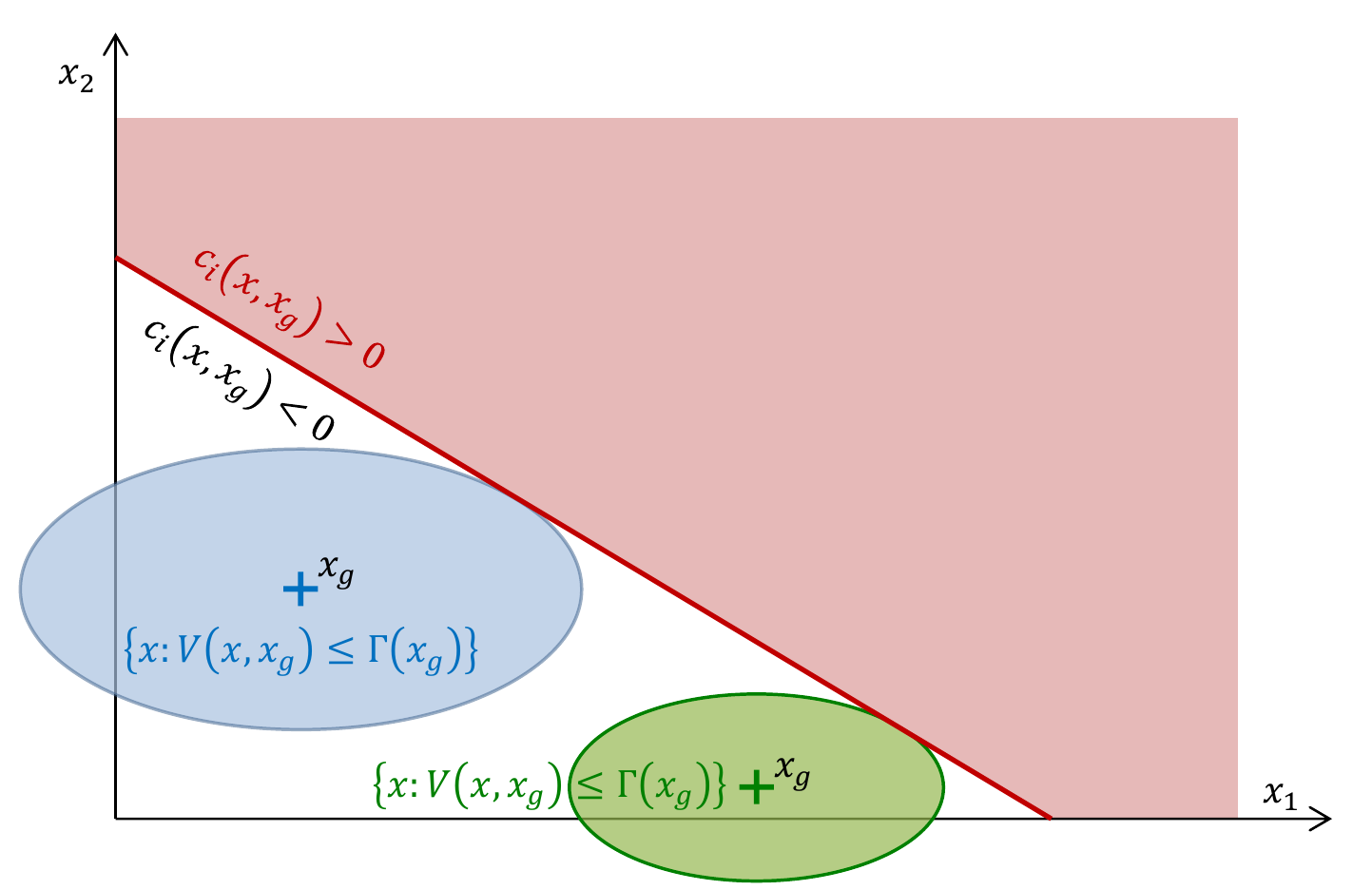}
{\caption{Geometrical interpretation of the $\Gamma_i(x_g)$ for two values of $g$} }
\end{figure}

At this point, given the currently applied $g(t)$ satisfying (\ref{LyapConstraints}) and given a desired reference at time $t,$ $r(t),$ the idea of the proposed Explicit Reference Governor (E-RG) is to keep enforcing (\ref{LyapConstraints}) in the future while manipulating the derivative $\dot{g}(t)$ so that $g(t)$ tends to $r(t)$.
 \\

To do so, the derivative $\dot{g}(t)$ is decomposed in its direction and module
\begin{equation}\label{RGFormula}
\dot{g}(t)=\hat{\rho}(t)\gamma(t).
\end{equation}
The direction $\hat{\rho}$ is chosen along the line connecting $g(t)$ and $r(t),$ i.e.
\begin{equation}\label{rho}
\hat{\rho}(t)=\frac{r(t)-g(t)}{\left\Vert r(t)-g(t)\right\Vert }.
\end{equation}
The amplitude $\gamma(t) \geq 0$ can be selected so that (\ref{LyapConstraints}) is satisfied, i.e.
    \begin{itemize}
    \item if $g(t)=r(t),$ then $\gamma(t)=0$
    \item if $V(x(t),x_{g(t)})<\Gamma_{\mathcal{I}}(g(t))$ then $\gamma(t)$ can be an arbitrarily high positive value
    \item if $V(x(t),x_{g(t)})=\Gamma_{\mathcal{I}}(g(t))$ then $\gamma(t)$ must be such that $\dot{V}\left(x,x_{g},\dot{g}(t)) \right)\leq \underset{i\in\mathcal{I}}{\min}\left(\dot{\Gamma}_{i}\left(x_{g},\dot{g}(t)\right)\right).$
    \end{itemize}
\vspace{0.5 cm}
In this paper, the following law for $\gamma(t)=\gamma(x(t),g(t),r(t))$ is proposed
\begin{equation}\label{FormulaRG}
\gamma(x,g,r)=\left[\nu\left(x,g,r\right)+\phi\left(x,g\right)\right]\sigma\left(g,r\right)l\left(g,r\right)
\end{equation}
where
\begin{enumerate}
\item $\nu\left(x,g,r\right)$ is a finite and non-negative feedforward term such that
\begin{equation}\label{nuCondition}
\dot{V}\left(x,x_{g},\hat{\rho} \mu)\right) \leq \underset{i\in\mathcal{I}}{\min}\left(\dot{\Gamma}_{i}\left(x_{g},\hat{\rho} \mu) \right)\right),\,\,\, \forall\mu:\, 0\leq\nu\leq\nu(x,g,r)
\end{equation}
\item $\phi(x,g)$ is a non-negative feedback term based on the distance between $\Gamma_{\mathcal{I}}(g)$ and $V(x,x_g):$
\begin{equation}\label{phiCondition}
    \phi(x,g)=\kappa\left(\Gamma_{\mathcal{I}}(x_g)-V(x,x_g)\right)
\end{equation}
     with $\kappa>0$ an arbitrary large scalar;
\item $\sigma\left(g,r\right)$ is a smoothing term introduced to smoothly stop $g$ when $g(t)=r(t)$
\begin{equation}\label{sigmaCondition}
\sigma=\min\left(1,\frac{\left|r-g\right|}{\epsilon_1}\right)
\end{equation}
with $\epsilon_1>0$ an arbitrary small scalar.
\item $l(g,r)$ is a limiting term that prevents the steady state reference $x_g$ from exiting the constraints
\begin{equation}\label{hCondition}
l\left(g,r\right)=\left\{ \begin{array}{ll}
\min\left(1,\frac{\Gamma_{\mathcal{I}}\left(x_{g}\right)-\epsilon_{2}}{\epsilon_{2}}\right) & \mathrm{if}\;\underset{i\in\mathcal{I}}{\min}\left(\nabla_g\Gamma_{i} \cdot \hat{\rho}\right)<0\\
1 & \mathrm{if}\;\underset{i\in\mathcal{I}}{\min}\left(\nabla_g\Gamma_{i} \cdot \hat{\rho}\right)\geq0
\end{array}\right.
\end{equation}
where $\epsilon_2>0$ is an arbitrary small scalar. With a slight abuse of notation, $\nabla_{g} \Gamma_{i}$ denotes the gradient vector evaluated in $g,$ i.e.
\begin{equation}
\nabla_g\Gamma=\left.\left[\frac{\partial\Gamma(x_g)}{\partial g_{1}},...,\frac{\partial\Gamma(x_g)}{\partial g_{m}}\right]\right|_{{g}}.
\end{equation}
\end{enumerate}
The structure of the proposed Explicit Reference Governor is depicted in Figure 2. The following result can be proven
\begin{figure}
\label{ERGscheme}
\centering
\includegraphics[width=9cm]{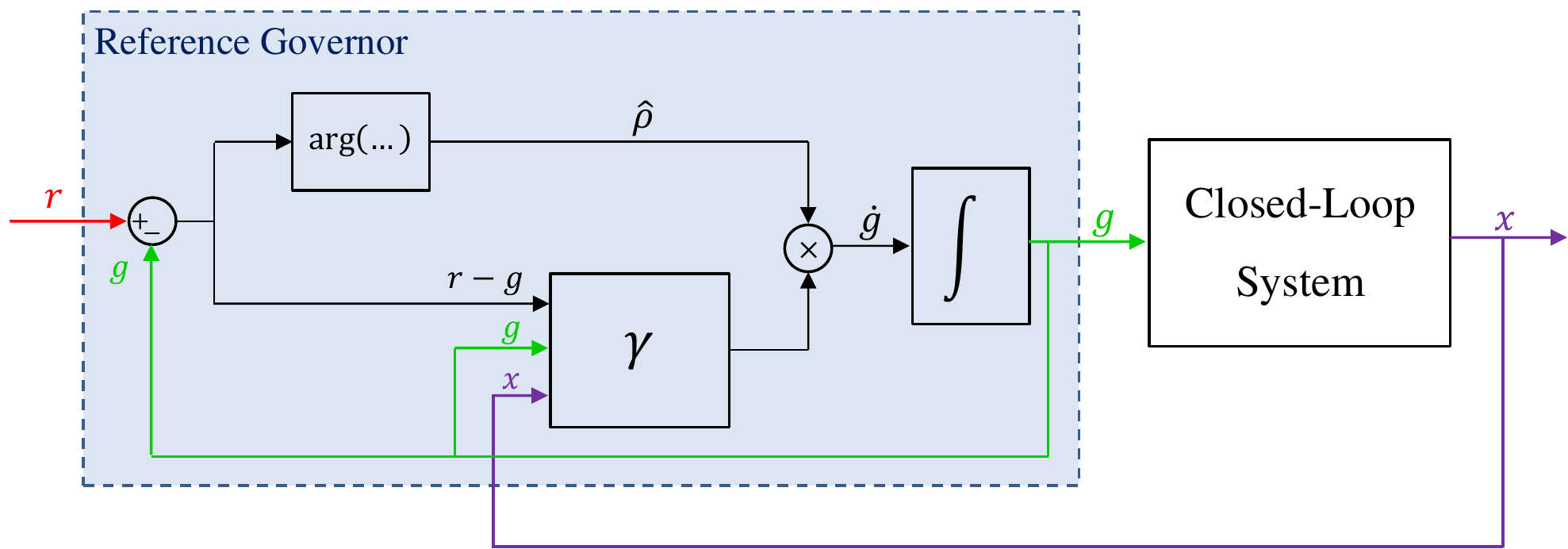}
\vspace{-0.3cm}
{\caption{Explicit Reference Governor Scheme} }
\end{figure}
\begin{teorema}
Let (\ref{system}) be a system subject to constraints (\ref{constraints}) and let $V(x,x_g)$ be a Lyapunov function such that, for any $g$,
the corresponding equilibrium $x_g \in \Real^n$ is proven to be GAS. Moreover, let functions $\Gamma_{i}\left(x_{g}\right), i=1,...,n_c$ exist such that (\ref{conditionOnGammai}) holds true. If the applied reference $g(t)$ is changed accordingly with (\ref{rho})-(\ref{hCondition}) and if, at at time $t=0$, a reference $g(0)$ is applied such that $V(x(0),x_{g(0)}) < \Gamma_{\mathcal{I}}(g(0))+\epsilon_2$ then:
\begin{itemize}
\item[a)] if $r(t)$ is piecewise continuous, $g(t)$ is continuous;
\item[b)] for any piecewise continuous signal $r(t)$, constraints (\ref{constraints}) are never violated;
\item[c)] if $r(t)=\bar{r}$ is kept constant form $t=0$ onward, then $g(t)$ asymptotically tends to $\bar{r}$ if $\Gamma_{\mathcal{I}}(\bar{r})\geq\epsilon_2.$ Otherwise, $r(t)$ tends to $r^*,$ which is the best approximation of $\bar{r}$ along the direction $\hat{\rho}(0)$ such that $\Gamma_{\mathcal{I}}(r^*)\geq\epsilon_2.$
\end{itemize}
\end{teorema}%
{\it Proof }
\begin{itemize}
\item[a)] Consider the Ordinary Differential Equations characterizing the system with the proposed reference governor
\[
\left[\!
  \begin{array}{c}
    \dot{x} \\
    \dot{g} \\
  \end{array}\!
\right]\!=\!\left[\!
  \begin{array}{c}
    f(x(t),g(t)) \\
    \hat{\rho}(g(t),r(t)) \gamma(x(t),g(t),r(t)) \\
  \end{array}\!
\right]\!=\!{\cal F}(x(t),g(t),r(t)).
\]
Given the definitions of (\ref{rho}) and (\ref{FormulaRG}), at each time instant $t$ (and so for a fixed $r(t)$), ${\cal F}(x,g,r)$ is Lipshitz continuous for any $x \in \Real^n$ and for any $g : \Gamma_{\mathcal{I}}(g) > \epsilon_2.$ Moreover, for whatever piecewise continuous $r(t),$
${\cal F}(x,g,r(t))$ is piecewise continuous since its discontinuities may arise only as a consequence of the countable discontinuities in $r(t)$ or as an effect of the countable times that $r(t)$ crosses $\Gamma_{\mathcal{I}}(r) = \epsilon_2.$ Using the fundamental theorem of Ordinary Differential Equations \cite{Callier}, this implies that, given  initial conditions $x(0),g(0)$, a solution $x(t)$ and $g(t)$ exists, is continuous and is unique for any finite $t$ such that $\Gamma_{\mathcal{I}}(g(\tau)) > \epsilon_2, \forall \tau \in [0,t]$. \\
To complete the proof, it is necessary to show that for any initial conditions such that  $V(x(0),x_{g(0)}) < \Gamma_{\mathcal{I}}(g(0))+\epsilon_2,$ then $\Gamma_{\mathcal{I}}(g(t))> \epsilon_2$ for all finite $t.$ This invariance property is guaranteed by the term (\ref{hCondition}) which ensures that $\lim_{\Gamma_{\mathcal{I}}(g(\tau))\rightarrow \epsilon_2} \dot{g}=0.$

\item[b)]By the definition of $\Gamma_{\mathcal{I}}$ in (\ref{conditionOnGammai}), the constraints are satisfied whenever condition (\ref{LyapConstraints}) is satisfied.
    Since functions $V\left(x,x_{g}\right)$, $\Gamma_{\mathcal{I}}\left(x_{g}\right)$, $x_{g}$, $g(t)$ and $x(t)$ are continuous, $\Gamma_{\mathcal{I}}(t)-V(t)$ is continuous, as well. This implies that if $V(t_1)<\Gamma_{\mathcal{I}}(t_1)$ at a certain time $t_1,$ then the only possibility to have $V(t_3)>\Gamma_{\mathcal{I}}(t_3)$ in a future time $t_3>t_1$ is to pass at time $t_2\in(t_1,t_3)$ through $V(t_2)-\Gamma_{\mathcal{I}}(t_2)=0$. However, whenever $V(x,x_g)=\Gamma_{\mathcal{I}}(x_g),$ the feedback term is equal to $\phi=0.$ Moreover since $\sigma(g,r)\leq 1 ,l(g,r) \leq 1$ it follows by definition that
    \[
    \gamma(x,g,r)=\nu\left(x,g,r\right)\sigma\left(g,r\right)l\left(g,r\right)\leq\nu\left(x,g,r\right).
    \]
    Due to equation (\ref{nuCondition}), this implies $\dot{V}\left(t_2 \right) \leq \underset{i\in\mathcal{I}}{\min}\left(\dot{\Gamma}_{i}\left(t_2 \right)\right)$, which prevents $V(t)$ from being greater than $\Gamma_{\mathcal{I}}(t)$.
\item[c)]Define the reference offset $\tilde{g}=r-g$. For $r(t)=\bar{r}$, it follows that $\dot{\tilde{g}}=\dot{g}=\hat{\rho}\gamma$ can be re-written as
    \[
    \dot{\tilde{g}}=-\frac{\tilde{g}}{\left\Vert\tilde{g}\right\Vert}\gamma(t)
    \]
    where $\gamma(t)$ is a nonnegative external input. Consider the Lyapunov function $\mathcal{V}=\frac{1}{2}\tilde{g}^2$. Its derivative is
    \[
    \dot{\mathcal{V}}=-\gamma(t)\left\Vert\tilde{g}\right\Vert
    \]
    which (being $\gamma(t) \geq 0$ by construction) is negative definite $\forall t:\gamma(t)\neq 0$. It is therefore enough to study under which conditions $\gamma(t)$ remains zero. Following from (\ref{FormulaRG}), $\gamma(t)$ is zero if
    \begin{itemize}
    \item $\sigma\left(g,r\right)=0$. This is possible only if $g=r$, i.e. the applied reference has converged to the desired one.
    \item $\nu(x,g,r)+\phi\left(x,g\right)=0$. This is true only if $\nu(x,g,r)=0$ and $V\left(x,x_{g}\right)=\Gamma_{\mathcal{I}}\left(x_{g}\right)$. Assume this happens at a certain time $t_S$. In this case, the applied reference remains constant and, due to the asymptotic stability of the system, it follows that $\exists\,t_M>t_S:\quad \dot{V}(t_M)<0$. Therefore, at time $t_M$ or just after it, either the feedforward term $\nu\left(x,g,r\right)$ and/or the feedback term $\phi\left(x,g\right)$ will no longer be equal to zero.
    \item $l(g,r)=0$. This term is the only one that can become (and remain indefinitely) zero for $g\neq r$. Please note that this happens only when $\Gamma_{\mathcal{I}}\left(x_{g}\right)=\epsilon_2$ and $\hat{\rho}$ points towards the outside of the set $\Gamma_{\mathcal{I}}(x_g)>\epsilon_2$. In this case, the applied reference is a constant $g(t)=r^*$. Although this implies $g \neq \bar{r}$, the applied reference cannot proceed any further in the direction $\hat{\rho}$ without violating the safety margin $\Gamma_\mathcal{I}(x_g)\geq\epsilon_2$. Since the constraints are convex, $r^*$ is the best approximation with margin $\varepsilon_2$ along the direction $\hat{\rho}$
    \end{itemize}
\end{itemize}
$\hfill\square$
\begin{remark}\label{RAS} The results presented in this paper also hold true for Regional Asymptotic Stability under the condition that an inner approximation of the attraction basin centered in $x_g$ is known for each $g$. Indeed, in this case it is enough to enforce as an additional constraint that the state is contained in the attraction basin.$\hfill \Box$
\end{remark}
\vspace{12 pt}
The two main challenges to make use of the proposed reference governor consist of:
\begin{itemize}
\item computing functions $\Gamma_{i}(x_g), i=1,...,n_c$ such that (\ref{conditionOnGammai}) hold true;
\item computing (possibly in closed form) $\nu(x,g,r)$ such that (\ref{nuCondition}) holds true
\end{itemize}
Note that, although the computation of the maximal $\Gamma_{i}(x_g)$ and $\nu(x,g,r)$ can speed up performances, it may be cumbersome from a practical viewpoint. However, being (\ref{conditionOnGammai}) and (\ref{nuCondition}) inequalities, simpler $\Gamma_{i}(x_g)$ and $\nu(x,g,r)$ may be obtained by using suitable bounds of the Lyapunov function. Additionally, it worth noting that the expression $\nu(t)=0$, although conservative in line of principle, always satisfies (\ref{nuCondition}) and can be used in the absence of a better suited solution.

Section \ref{S2} will provide the closed-form expression of $\Gamma_{i}(x_g)$ for the large and highly relevant class of nonlinear systems subject to polyhedric constraints and admitting Lyapunov functions lower-bounded by quadratic forms.

\section{(Non-)Quadratic Lyapunov Function / Polytopic Constraints}\label{S2}
This section will show how to implement the E-RG in the case of a pre-stabilized system (\ref{system}) subject to polytopic constraints
\begin{equation}\label{polytopes}
c_{xi}^{T}x(t)+c_{gi}^{T}x_{g(t)}+d_{i}\geq 0, \forall t, \,\,\, i=1,...,n_c
\end{equation}
and whose Lyapunov function is (or is lower-bounded by) a quadratic form. This class is deemed highly relevant since any convex set can be approximated by a convex polyhedron and many physical systems present such Lyapunov functions.
\\

The following proposition provides the optimal $\Gamma_{i}(x_g)$ for quadratic Lyapunov functions.
\\

\begin{proposition} \label{prop: QuadLyap}
Consider the Lyapunov function
\begin{equation}\label{quadratic}
V\left(x,x_{g}\right)=\left(x-x_{g}\right)^{T}P\left(x-x_{g}\right)
\end{equation}
with $P>0.$ Under the condition $(c_{xi}+c_{gi})^T x_{g}+d_{i} \geq 0$, the largest $\Gamma_i(x_g)$ such that $V(x,x_g) \leq \Gamma_{i}(x_g)$ implies $c_{xi}^{T}x+c_{gi}^{T}x_{g}+d_{i}\geq 0$ is
\begin{equation}\label{eq: Gamma}
\Gamma_{i}(x_g)=
\begin{array}{ll}
\frac{\left((c_{xi}+c_{gi})^T x_{g}+d_{i}\right)^{2}}{c_{xi}^T P^{-1}c_{xi}}
\end{array}
\end{equation}
\end{proposition}
{\it Proof: }
Without loss of generality, the change of coordinates $\tilde{x}=P^{1/2}\left(x-x_{g}\right)$ is applied. The Lyapunov function (\ref{quadratic}) becomes
\begin{equation}
V\left(\tilde{x},x_{g}\right)=\tilde{x}^{T}\tilde{x}\label{eq: LyapunovSphere}
\end{equation}
which is a sphere centered in $x=x_{g}$. The constraint becomes
\[
c_{xi}^{T}P^{-1/2}\tilde{x}+(c_{xi}+c_{gi})^T x_{g}+d_{i}\geq 0.
\]
Each constraint boundary is then given by the hyperplane
\begin{equation}
\Pi_{i}\tilde{x}+\Delta_{i}=0\label{eq: Hyperplane}
\end{equation}
where
$\Pi_{i}^{T}=c_{xi}^{T}\sqrt{P}^{-1}$ and $\Delta_{i}=(c_{xi}+c_{gi})^T x_{g}+d_{i}.$
Therefore, the maximum value of the Lyapunov function is the square of the distance between
this hyperplane and the origin,
\[
\left|\tilde{x}\right|^2=\left(\frac{\left|\Delta_{i}\right|}{\left\Vert \Pi_{i}\right\Vert }\right)^2
\]
which implies $\Gamma_{i}=\tilde{x}^{T}\tilde{x}=\frac{1}{2}\left|\tilde{x}\right|^{2}.$
$\hfill \Box$
\\

\begin{remark}
Note that (\ref{eq: Gamma}) is defined only for command $g$ such that $(c_{xi}+c_{gi})^T x_{g}+d_{i} > 0,$ which are the only commands that can be selected by the E-RG. The case $(c_{xi}+c_{gi})^T x_{g}+d_{i} < 0$  can be covered by choosing whatever arbitrary negative $\Gamma(x_g)$ without changing in any way the behavior of the E-RG.
\end{remark}
\vspace{0.4 cm}

In the case of non-quadratic Lyapunov functions, the following proposition proves that equation (\ref{eq: Gamma}) provides a feasible $\Gamma_{i}(x_g),$ although it is no longer optimal.
\\

\begin{proposition}\label{prop: NonQuadLyap}
Let a Lyapunov function $V(x,x_g)$ be lower bounded by a quadratic function
\[
\underline{V}(x,x_g)= (x-x_g)^T P (x-x_g)\leq V(x,x_g), \forall x \in \Real^n, g \in \Real^m.
\]
The function $\Gamma_{i}(x_g)$ as in (\ref{eq: Gamma}) is such that $V(x,x_g) \leq \Gamma_{i}(x_g)$ implies $c_{xi}^{T}x+c_{gi}^{T}x_{g}+d_{i}\geq 0.$
\end{proposition}
{\it Proof: }
Following from Proposition \ref{prop: QuadLyap}, $\Gamma_{i}(x_g)$ as in (\ref{eq: Gamma}) is the largest function such that $\underline{V}(x,x_g) \leq \Gamma_{i}(x_g)$ implies $c_{xi}^{T}x+c_{gi}^{T}x_{g}+d_{i}\leq 0.$ Since $\underline{V}(x,x_g)\leq V(x,x_g)$, it follows that $V(x,x_g) \leq \Gamma_{i}(x_g)$ ensures the constraint satisfaction, as well. $\hfill \Box$
\\

In conclusion, for any nonlinear system belonging to this class, the E-RG strategy (\ref{rho})-(\ref{FormulaRG}) can be implemented by using:
\begin{itemize}
\item No feedforward term, i.e. $\nu\left(x,g,r\right)=0$;
\item the feedback term $\phi$ as in (\ref{phiCondition}) where $\Gamma_i(x_g)$ is given by (\ref{eq: Gamma});
\item $\sigma$ as in (\ref{sigmaCondition});
\item $l$ that can be simplified w.r.t. (\ref{hCondition}) as
\begin{equation}\label{hCondition2}
\begin{array}{ll}
l\left(g,r\right)=\\
\left\{\! \begin{array}{ll}
\!\min\left(\!1,\frac{\Gamma_{\mathcal{I}}\left(x_{g}\right)-\epsilon_{2}}{\epsilon_{2}}\!\right) & \mathrm{if}\,\underset{i\in\mathcal{I}}{\min}\left(\!(c_{xi}+c_{gi})^T \nabla_g x_{g}\cdot \hat{\rho}\!\right)\!<\!0\\
1 & \mathrm{if}\,\underset{i\in\mathcal{I}}{\min}\left(\!(c_{xi}+c_{gi})^T \nabla_g x_{g}\cdot \hat{\rho}\!\right)\!\geq\!0.
\end{array}\right.
\end{array}
\end{equation}
\end{itemize}
due to the fact that
\begin{equation}\label{eq: NablaGamma}
\nabla\Gamma_{i}=2\frac{(c_{xi}+c_{gi})^T x_g + d_i}{c_x^T P^{-1}c_x} (c_{xi}+c_{gi})^T\nabla_g x_{g}
\end{equation}
where $\nabla_g x_{g}$ only depends on the system equations at steady-state.
\begin{remark} The choice $\nu=0$ is due to the fact that, in the general case, the determination of a nonzero $\nu(x,g,r)$ in closed form could be prohibitive. However, as will be clearer in the numerical simulations, the effect of this term on the E-RG performance is usually negligible for a high enough $\kappa$ in the feedback term $\phi$. The two following sections specialize the very general results of this section to two notable cases where $\nu$ is computable: Linear Systems and Robotic Manipulators.
\end{remark}

\section{Linear Systems}\label{sec: Linear Systems}
A notable case of systems that falls into the class of the systems described in Section \ref{S2} are linear systems subject to polytopic constraints. Interestingly enough, for linear systems it is possible to compute not only the maximal $\Gamma_i(x_g)$, but also the maximal $\nu(x,g,r).$

Consider a linear system
\begin{equation}\label{linsys}
\dot{x}=Ax(t)+Bg(t)
\end{equation}
where $A$ is Hurwitz and subject to constraints (\ref{polytopes}). Given a constant $g$, the associated steady state is $x_{g}=-A^{-1}Bg.$
Moreover, being $A$ asymptotically stable, for any $Q>0$ a $P>0$ exists such that
\begin{equation}\label{Lyapeq}
A^T P + P A = -Q.
\end{equation}
\begin{proposition} Consider the GAS linear system (\ref{linsys}) subject to constraints (\ref{polytopes}) and let $P>0,Q>0$ satisfy the Lyapunov equation (\ref{Lyapeq}). Given $\Gamma_i(x_g)$ as defined in (\ref{eq: Gamma}), then
\begin{equation}\label{eq: nulin}
\begin{array}{ll}
\nu(x,x_g)=\\
max\left\{0,\underset{i\in\mathcal{I}}{\min} \frac{\frac{1}{2}\left(x-x_{g}\right)^{T}Q\left(x-x_{g}\right)}{\left[\left(x-x_{g}\right)^{T}P+\frac{(c_{xi}+c_{gi})^{T}x_{g}+d_{i}}{c_{xi}^T P^{-1}c_{xi}}(c_{xi}+c_{gi})^{T}\right]A^{-1}B\hat{\rho}}\right\}
\end{array}
\end{equation}
is the largest $\nu$ satisfying (\ref{nuCondition}).
\end{proposition}
{\it Proof:}
Consider the Lyapunov function $V=\left(x-x_{g}\right)^{T}P\left(x-x_{g}\right).$ For a nonconstant $x_{g},$ the derivative is
\[
\begin{array}{l}
\dot{V}(x,x_g,\dot{x}_g)\!=\!-\!\left(\!x-x_{g}\!\right)^{T}\!Q\left(\!x-x_{g}\!\right)\!-\!\left[\!\left(\!x-x_{g}\!\right)^{T}\!P\dot{x}_{g}\!+\!\dot{x}_{g}^{T}\!P\left(\!x-x_{g}\!\right)\!\right].
\end{array}
\]
By substituting  $\dot{x}_{g}=-A^{-1}B\hat{\rho}\nu$, it follows that
\begin{equation}\label{Vdotlin}
\begin{array}{ll}
\dot{V}(x,x_g,\hat{\rho}\nu)=\\
-\!\left(\!x-x_{g}\!\right)^{T}\!Q\left(\!x-x_{g}\!\right)+\left[\!\left(\!x-x_{g}\!\right)^{T}\!PA^{-1}\!B\hat{\rho}+\hat{\rho}^{T}\!B^{T}\!A^{-T}\!P\left(\!x-x_{g}\!\right)\!\right]\nu\\
=-\left(x-x_{g}\right)^{T}Q\left(x-x_{g}\right)+ \left[2\left(x-x_{g}\right)^{T}PA^{-1}B\hat{\rho}\right]\nu.
\end{array}
\end{equation}
Likewise, for each constraint (\ref{eq: Gamma}), the time derivative is
\[
\dot{\Gamma}_{i}(x_g,\dot{g})=2\frac{(c_{xi}+c_{gi})^T x_{g}+d_{i}}{c_{xi}^T P^{-1}c_{xi}}(c_{xi}+c_{gi})^T\dot{x}_g.
\]
Again, by substituting $\dot{x}_{g}=-A^{-1}B\hat{\rho}\nu$, it follows that
\begin{equation}\label{dotGammai}
\dot{\Gamma}_{i}(x_g,\hat{\rho}\nu)=-2\frac{(c_{xi}+c_{gi})^T x_{g}+d_{i}}{c_{xi}^T P^{-1}c_{xi}}(c_{xi}+c_{gi})^TA^{-1}B\hat{\rho}\nu.
\end{equation}
Being both $\dot{V}$ and $\dot{\Gamma}_{i}$ affine in $\nu,$ the largest $\nu$ such that (\ref{nuCondition}) holds true is the one such that
\[
\dot{V}\left(x,x_{g},\hat{\rho} \nu)\right) = \underset{i\in\mathcal{I}}{\min}\left(\dot{\Gamma}_{i}\left(x_{g},\hat{\rho} \nu) \right)\right)
\]
which is (\ref{eq: nulin}).
$\hfill \Box$

In line of principle, the feedforward term (\ref{eq: nulin}) improves the convergence speed of $g(t).$ However, as shown in Section \ref{NumEx}, in many cases the improvement w.r.t. $\nu=0$ is marginal.

\begin{remark}
Note that in this case $\nabla x_{g}$ to be used in (\ref{hCondition}) is simply
\begin{equation}\label{linNablaxg}
\nabla x_{g}=-A^{-1}B.
\end{equation}
\end{remark}

\begin{remark}
Note that since for any $Q>0$ there exists a $P>0$ satisfying (\ref{Lyapeq}), $Q$ can be used as a degree of freedom to tune the E-RG performances.
\end{remark}

\section{Robotic Systems}
This section specializes the procedure detailed in Section \ref{S2} to robotic systems controlled with a PD with gravity compensation and subject to polytopic constraints (\ref{polytopes}). Interestingly enough, for this class of systems Proposition \ref{prop: NonQuadLyap} applies and, moreover, a simple closed-form expression for the feedforward term $\nu(x,g,r)$ can be provided.

Consider a robotic system described by the dynamic equations
\begin{equation}\label{Robot}
M\left(q\right)\ddot{q}+C\left(q,\dot{q}\right)\dot{q}+G\left(q\right)=u
\end{equation}
controlled with a standard PD with gravity compensation
\begin{equation}\label{Pdgrav}
u=G\left(q\right)-K_{P}\left(q-g\right)-K_{D}\dot{q},
\end{equation}
where $K_{P}>0$, $K_{D}>0$ are diagonal matrices. By defining the state as $x=[q^T,\dot{q}^T]^T$, is is well-known that any equilibrium point
\begin{equation}\label{xgRobot}
x_g=\left[\begin{array}{c}
g\\
0
\end{array}\right]
\end{equation}
is proven to be GAS using the LaSalle Criterion and the Lyapunov function
\begin{equation}\label{robotLyap}
V=\frac{1}{2}
(x-x_g)^T
\left[
                          \begin{array}{cc}
                            K_{P} & 0 \\
                            0 & M\left(q\right) \\
                          \end{array}
                        \right](x-x_g).
\end{equation}
Moreover, due to the properties of the mass matrix $M(q),$ it is always possible to compute a matrix $\underline{M}>0$ such that $M(q)-\underline{M} \geq 0, \forall q$. As a result, by choosing
\begin{equation}\label{eq: Probot}
P=\frac{1}{2}\left[\begin{array}{cc}
                    K_{P} & 0 \\
                    0 & \underline{M} \\
                    \end{array}\right],
\end{equation}
functions $\Gamma_i(x_g)$ can be computed as in (\ref{eq: Gamma}). The following proposition provides the closed form expression of $\nu(x,g,r).$
\begin{proposition}
Let the system (\ref{Robot})-(\ref{Pdgrav}) be subject to constraints (\ref{polytopes}) and let (\ref{robotLyap}) be the Lyapunov function used to prove GAS using the LaSalle Criterion. Defining $P>0$ as in (\ref{eq: Probot}) and using $\Gamma_i(x_g)$ as defined in (\ref{eq: Gamma}), the feedforward term
\begin{equation}\label{nurobot}
\nu(x,g,r)=max\left\{0,\underset{i\in\mathcal{I}}{\min} \frac {\dot{q}^{T}K_{D}\dot{q}}{2\frac{c_{Ri}^T g+d_{i}}{c_{xi}^T P^{-1}c_{xi}}c_{Ri}^T \hat{\rho}-\left(q-g\right)^{T}K_{P}\hat{\rho}}\right\}
\end{equation}
with $c_{Ri}^T=\left[c_{xi}+c_{gi}\right]^T\left[I_n,0_{n \times n}\right],$ is the largest one ensuring condition (\ref{nuCondition}).
\end{proposition}
{\it Proof:} \\
In the presence of a time varying reference, the derivative of (\ref{robotLyap}) is
\begin{equation}\label{dotVrob}
\dot{V}(x,x_g,\dot{g})=-\dot{q}^{T}K_{D}\dot{q}-\left(q-g\right)^{T}K_{P}\dot{g}.
\end{equation}
By substituting $\dot{g}=\hat{\rho}\nu$ the latter becomes
\begin{equation}\label{dotVrob}
\dot{V}(x,x_g,\hat{\rho} \nu)=-\dot{q}^{T}K_{D}\dot{q}-\left(q-g\right)^{T}K_{P}\hat{\rho} \nu.
\end{equation}
The derivative of $\Gamma_i(x_g)$ can be simplified as
\begin{equation}\label{dotGamma2}
\dot{\Gamma}_{i}(x_g,\hat{\rho}\nu)=-2\frac{(c_{xi}+c_{gi})^T  \left[\begin{array}{c}
                                                                                         g \\
                                                                                         0
                                                                                       \end{array}\right]+d_{i}}{c_{xi}^T P^{-1}c_{xi}}(c_{xi}+c_{gi})^T \left[\begin{array}{c}
                                                                                         \hat{\rho}\\
                                                                                         0
                                                                                       \end{array}\right]\nu.
\end{equation}
As in the linear case, being both $\dot{V}$ and $\dot{\Gamma}_{i}$ affine in $\nu,$ the largest $\nu$ such that (\ref{nuCondition}) holds true is the one such that $\dot{V}\left(x,x_{g},\hat{\rho} \nu)\right) = \underset{i\in\mathcal{I}}{\min}\left(\dot{\Gamma}_{i}\left(x_{g},\hat{\rho} \nu \right)\right),$ which is (\ref{nurobot}). $\hfill \Box$
\\

\begin{remark}
Note that, in this case, $\nabla x_{g}$ to be used in (\ref{hCondition}) is
\begin{equation}\label{NablaxgRobot}
\nabla x_g=\left[\begin{array}{c}
I_{n}\\
0_{n\times n}
\end{array}\right]
\end{equation}
\end{remark}

\section{Numerical Examples}\label{NumEx}
\subsection{Example: Linear Second Order System}
In this example the proposed E-RG will be compared with the classical prediction-based reference governor for discrete time linear systems introduced in \cite{Gilbert94}. Consider a double integrator
\[
\dot{x}(t)=\left[\begin{array}{cc}
0 & 1\\
0 & 0
\end{array}\right]x(t)+\left[\begin{array}{c}
0\\
1
\end{array}\right]u(t)
\]
controlled by a PD, i.e. $u=k_p(g - x_1)-k_d x_2,$ with $k_p,k_d>0$. The closed-loop expression of the controlled system is
\[
\dot{x}(t)=\left[\begin{array}{cc}
0 & 1\\
-k_{p} & -k_{d}
\end{array}\right]x(t)+\left[\begin{array}{c}
0\\
k_{p}
\end{array}\right]g(t)
\]
which presents GAS equilibria in $x_g=[g\,\,0]^T$ for all $g \in \Real$. It is assumed that the system is subject to state and input constraints
$
|x_1| \leq x_{max}$ and $|u| =\left|k_{p}\left(x-g\right)+k_{d}\dot{x}\right|\leq u_{max}$ corresponding to the affine constraints $c_{xi}^Tx+c_{gi}^Tx+d_i \geq 0, i=1,...,4$ where
\[
\begin{array}{lll}
c_{x1}=[1\,\,\, 0]^T, & c_{g1}=[0\,\,\, 0]^T, & d_1=x_{max}, \\
c_{x2}=[-1\,\,\, 0]^T, & c_{g2}=[0\,\,\, 0]^T, & d_2=x_{max}, \\
c_{x3}=[k_p\,\,\, k_d]^T, & c_{g3}=[-k_p\,\,\, 0]^T, & d_3=u_{max}, \\
c_{x4}=[-k_p\,\,\, -k_d]^T, & c_{g4}=[k_p\,\,\, 0]^T, & d_4=u_{max}.
\end{array}
\]
To implement the E-RG, functions $\Gamma_i(x_g)$ and $\nu\left(x,g,r\right)$ are defined as in (\ref{eq: Gamma}) and (\ref{eq: nulin}), respectively with
\[
P=\left[
     \begin{array}{cc}
       \frac{k_d}{k_p}+\frac{k_p}{k_d}+\frac{1}{k_d} & \frac{1}{k_p}  \\
       \frac{1}{k_p} & \frac{k_p+1}{k_d k_p} \\
     \end{array}
   \right],
\]
which is the solution of the Lyapunov equation $A^TP+PA=-Q$ for $Q=I_2.$
\begin{figure}
\label{fig: gLin1}
\centering
\includegraphics[width=8cm]{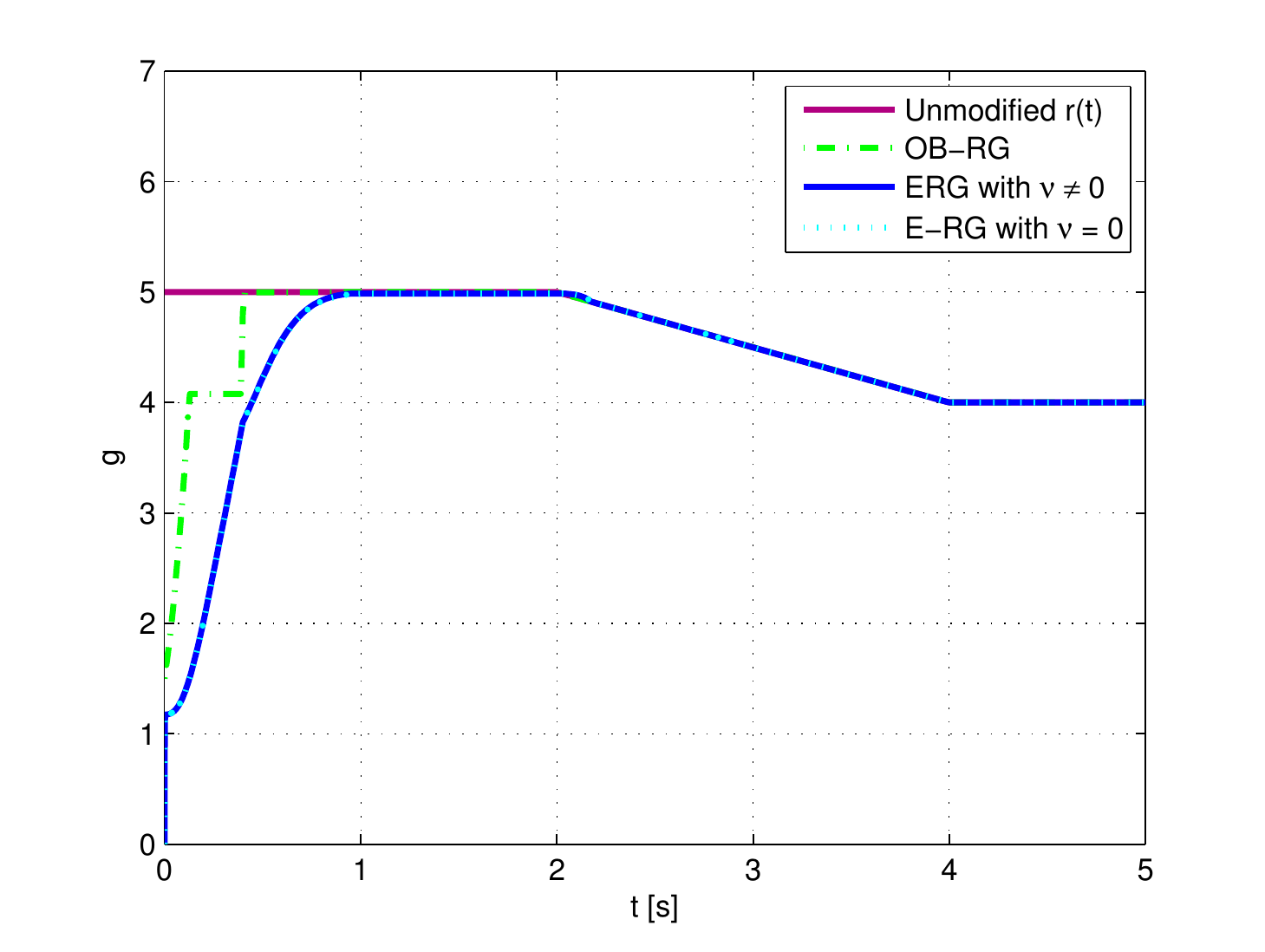}
\vspace{-0.3cm}
{\caption{Applied Reference - Linear System} }
\end{figure}
\begin{figure}
\label{fig: xLin1}
\centering
\includegraphics[width=8cm]{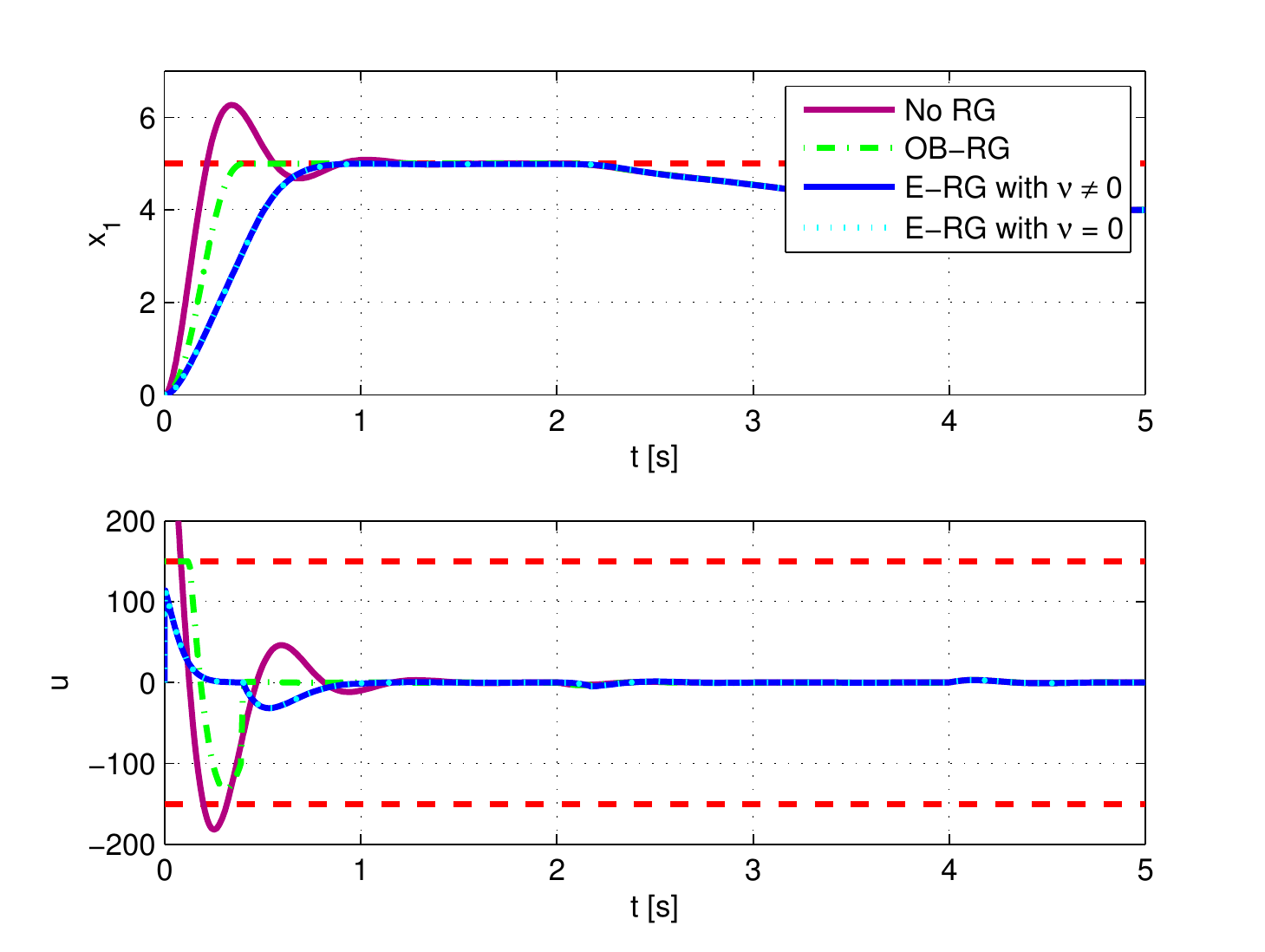}
\vspace{-0.3cm}
{\caption{Closed-Loop Response - Linear System} }
\end{figure}
Figures 3-4 show the simulations obtained for $k_p=100, k_d=8$. The following cases compared:
\begin{itemize}
\item \textbf{No RG:} The system is not provided with a Reference Governor and the desired reference is directly applied to the controlled system;
\item \textbf{Optimization-Based RG (OB-RG):} The Reference Governor based proposed in \cite{Gilbert94} is used. The system is sampled with a sampling time of $T_S=0.01 s$ and the constrained optimization problem is solved over a prediction horizon of $0.5 s;$
\item \textbf{Explicit RG, with Feedforward:} The E-RG is implemented using  $\epsilon_{1}=10^{-3}, \epsilon_{2}=10^{-3}$ and $\kappa=10^2$. The feedforward term given in (\ref{eq: nulin}) is used;
\item \textbf{Explicit RG, no Feedforward:} The E-RG is implemented with $\nu=0.$
\end{itemize}

Simulations show that all of the RG strategies are successful at enforcing the system constraints when necessary. As expected the E-RG strategy has a slower settling time than the OB-RG. This is due to the fact that the OB-RG is a model-predictive strategy whereas the E-RG does not explicitly use state predictions. However, as shown in Table \ref{TableTempi}, this loss in settling time is compensated by a drastic reduction of the computational time. This makes E-RG of interest for applications where real-time requirements and computational resources are incompatible with an optimization procedure. As for the feedforward term, note that the difference between the two E-RG strategies is marginal and justifies the omission of the feedforward term in the cases where a closed-form expression for $\nu(x,g,r)$ is difficult to obtain.
\begin{table}
\begin{center}
\begin{tabular}{|l|c|c|}
\hline
 & \multicolumn{1}{|c|}{E-CG} & {OB-RG} \\ \hline
{\small CPU Time [s/iteration]} & {0.0014} & {0.1617}\\
{\small Settling time [s]} & {0.61} & {0.31}\\
\hline
\end{tabular}
\end{center}
\caption{Computational time and settling time comparisons}
\label{TableTempi}
\end{table}
\subsection{Example: Two-Link Planar Arm}
\begin{figure}
\label{fig: RR_Arm}
\centering
\includegraphics[width=8cm]{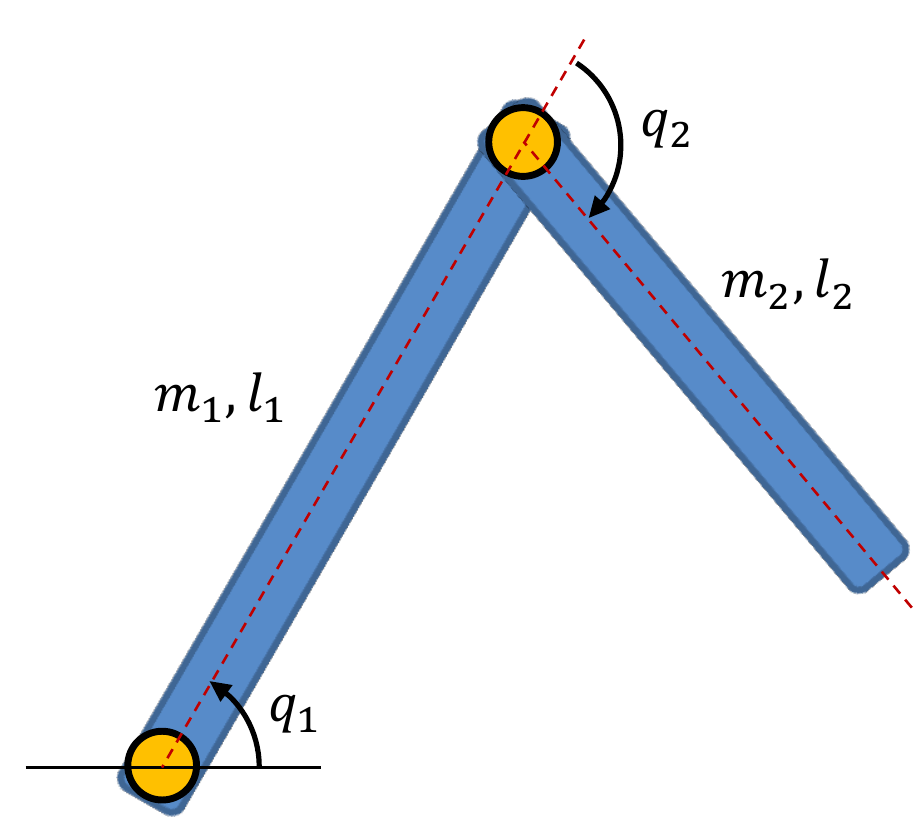}
\vspace{-0.3cm}
{\caption{Two-Link Planar Arm Model} }
\end{figure}
Consider the dynamic model of a planar arm with two rotational joints illustrated in Figure 5. The dynamics of the system is governed by equation (\ref{Robot}) with
\[
M\left(q\right)=\left[\begin{array}{cc}
\mu_{1}+2\nu\cos q_{2} & \mu_{2}+\nu\cos q_{2}\\
\mu_{2}+\nu\cos q_{2} & \mu_{2}
\end{array}\right]
\]
\[
C\left(q,\dot{q}\right)=\left[\begin{array}{cc}
-\nu\sin q_{2}\dot{q}_{2} & -\nu\sin q_{2}\left(\dot{q}_{1}+\dot{q}_{2}\right)\\
\nu\sin q_{2}\dot{q}_{1} & 0
\end{array}\right]
\]
\[
G\left(q\right)=a_{g}\left[\begin{array}{l}
\left(m_1 r_1+m_2 l_1\right)cos\left(q_1\right) + m_2 r_2cos\left(q_1+q_2\right))\\
m_2 r_2cos\left(q_1+q_2\right))\\
\end{array}\right]
\]
where
\[
\begin{array}{l}
\mu_{1}=\mathcal{I}_{z1}+\mathcal{I}_{z2}+m_{1}r_{1}^{2}+m_{2}\left(l_{1}^{2}+r_{2}^{2}\right)\\
\mu_{2}=\mathcal{I}_{z2}+m_{2}r_{2}^{2}\\
\nu=m_{2}l_{1}r_{2}.
\end{array}
\]
The physical parameters are $m_1=4 kg,\, m_2=3 kg, \, l_1=0.4m, \, l_2=0.3m$ and $r_i=l_i/2, \, \mathcal{I}_i=m_il_i^2/12,\,\, i=1,2$.

The robot is controlled using the PD with gravity compensation defined in (\ref{Pdgrav}) where $K_P = diag(k_{p1},k_{p2}), \, K_D = diag(k_{d1},k_{d2})$ with $k_{p1}=65$, $k_{p2}=45$, $k_{d1}=1.6$, $k_{d2}=1.3$, ensuring that each point of equilibrium (\ref{xgRobot}) is GAS for any constant reference $g=[g_1\,\,\,\, g_2]^T.$

It is assumed that the system is subject to the state and input constraints $q_1\in[q_{1min},q_{1max}],$ $q_2\in[q_{2min},q_{2max}],$ $\left|u_1\right|\leq u_{1max}$ and $\left|u_2\right|\leq u_{2max}$ with $q_{1min}=\frac{2\pi}{9}$, $q_{1max}=\frac{7\pi}{9}$, $q_{2min}=-\pi$, $q_{1max}=-\frac{\pi}{4}$, $u_{1max}=35 Nm$ and $u_{2max}=25 Nm.$\\
The state constraints becomes
\[
\begin{array}{lll}
c_{x1}=[\,\,\,\,\,1\,\,\, \,\,\,\,\,0 \,\,\, 0 \,\,\, 0]^T, & c_{g1}=0_{4\times1}, & d_1=-q_{1min}, \\
c_{x2}=[-1\,\,\, \,\,\,\,\,0 \,\,\, 0 \,\,\, 0]^T, & c_{g2}=0_{4\times1}, & d_2=\,\,\,\,\,q_{1max}, \\
c_{x3}=[\,\,\,\,\,0\,\,\, \,\,\,\,\,1 \,\,\, 0 \,\,\, 0]^T, & c_{g3}=0_{4\times1}, & d_3=-q_{2min}, \\
c_{x4}=[\,\,\,\,\,0\,\, -\!1 \,\,\, 0 \,\,\, 0]^T, & c_{g4}=0_{4\times1}, & d_4=\,\,\,\,\,q_{2max}.
\end{array}
\]
The input constraints converts into
\[
\left|-k_{Pi}\left(q_i-g_i\right)-k_{Di}\dot{q}_i+G_i\left(q_1,q_2\right)\right|\leq u_{imax}\quad i=1,2
\]
which are not in the desired form (\ref{polytopes}). To solve this issue, the gravity compensation terms are upper-bounded using
\[\begin{array}{lll}
G_{1max}=a_g\left(\left(m_1 r_1+m_2 l_1\right)cos\left(q_{1min}\right) + m_2 r_2cos\left(q_{1min}+q_{2max}\right)\right) \\
G_{2max}=a_gm_2 r_2cos\left(q_{1min}+q_{2max}\right) \\
\end{array}\]
which allows to express the constraints in a linear form
\[
\begin{array}{lll}
c_{x5}=[k_{p1}\,\,\, 0 \,\,\, k_{d1} \,\,\, 0]^T, & c_{g5}=[-k_{p1}\,\,\, 0 \,\,\,0 \,\,\, 0]^T,\\
c_{x6}=[-k_{p1}\,\,\, 0 \,\,\, -k_{d1} \,\,\, 0]^T, & c_{g6}=[k_{p1}\,\,\, 0 \,\,\,0 \,\,\, 0]^T,\\
c_{x7}=[0 \,\,\,k_{p2} \,\,\, 0 \,\,\, k_{d2}]^T, & c_{g7}=[0 \,\,\,-k_{p2} \,\,\,0 \,\,\, 0]^T,\\
c_{x8}=[0 \,\,\,-k_{p2} \,\,\, 0 \,\,\,-k_{d2}]^T, & c_{g8}=[0 \,\,\,k_{p2} \,\,\,0 \,\,\, 0]^T,\\
\\
d_5=u_{1max}-G_{1max}, \\
d_6=u_{1max}-G_{1max}, \\
d_7=u_{2max}-G_{2max}\\
d_8=u_{2max}-G_{2max}.
\end{array}
\]
\begin{figure}
\label{fig: xRbt}
\centering
\includegraphics[width=8cm]{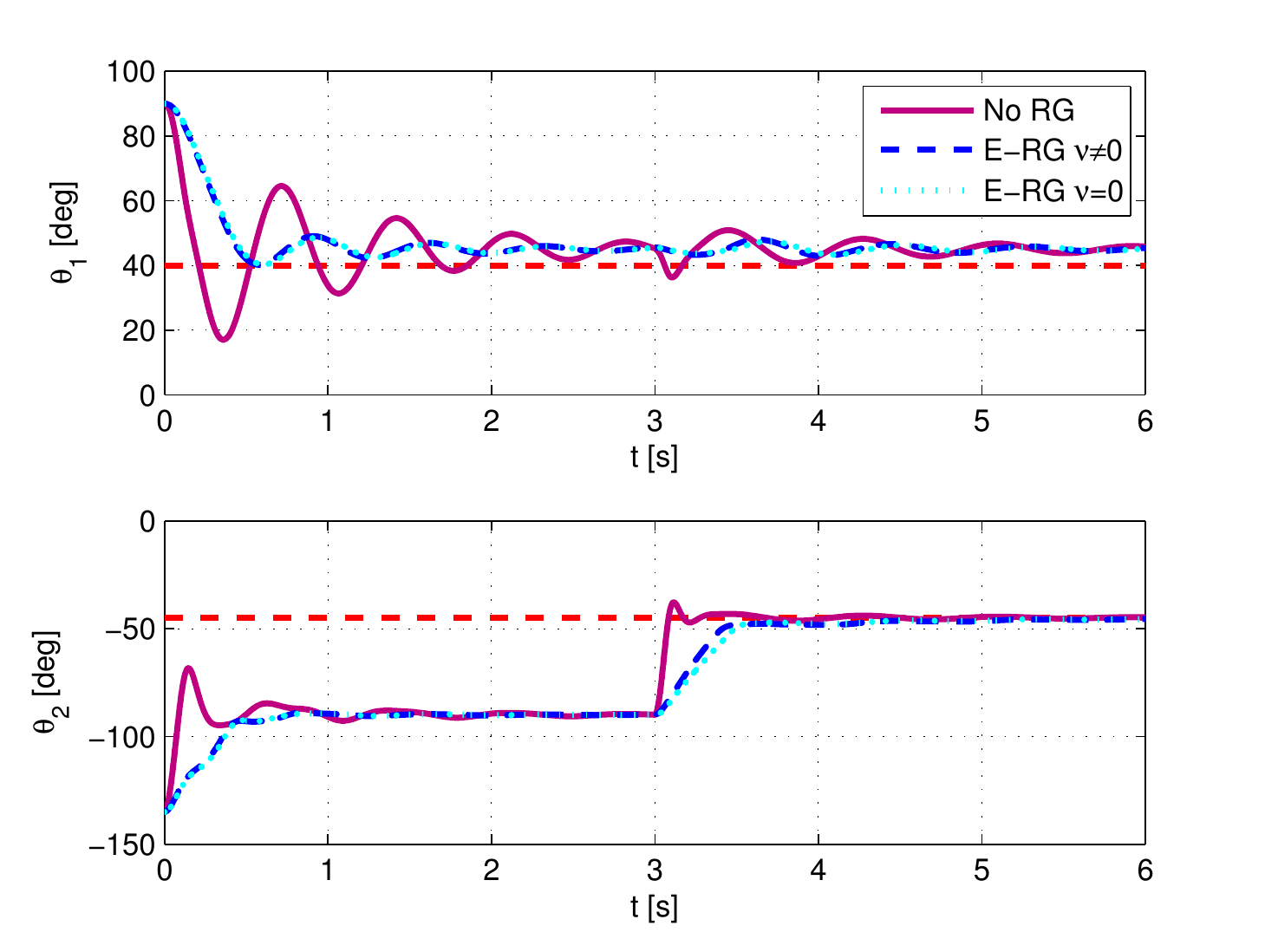}
\vspace{-0.3cm}
{\caption{Closed-Loop Response - Robotic Planar Arm} }
\end{figure}
\begin{figure}
\label{fig: uRbt}
\centering
\includegraphics[width=8cm]{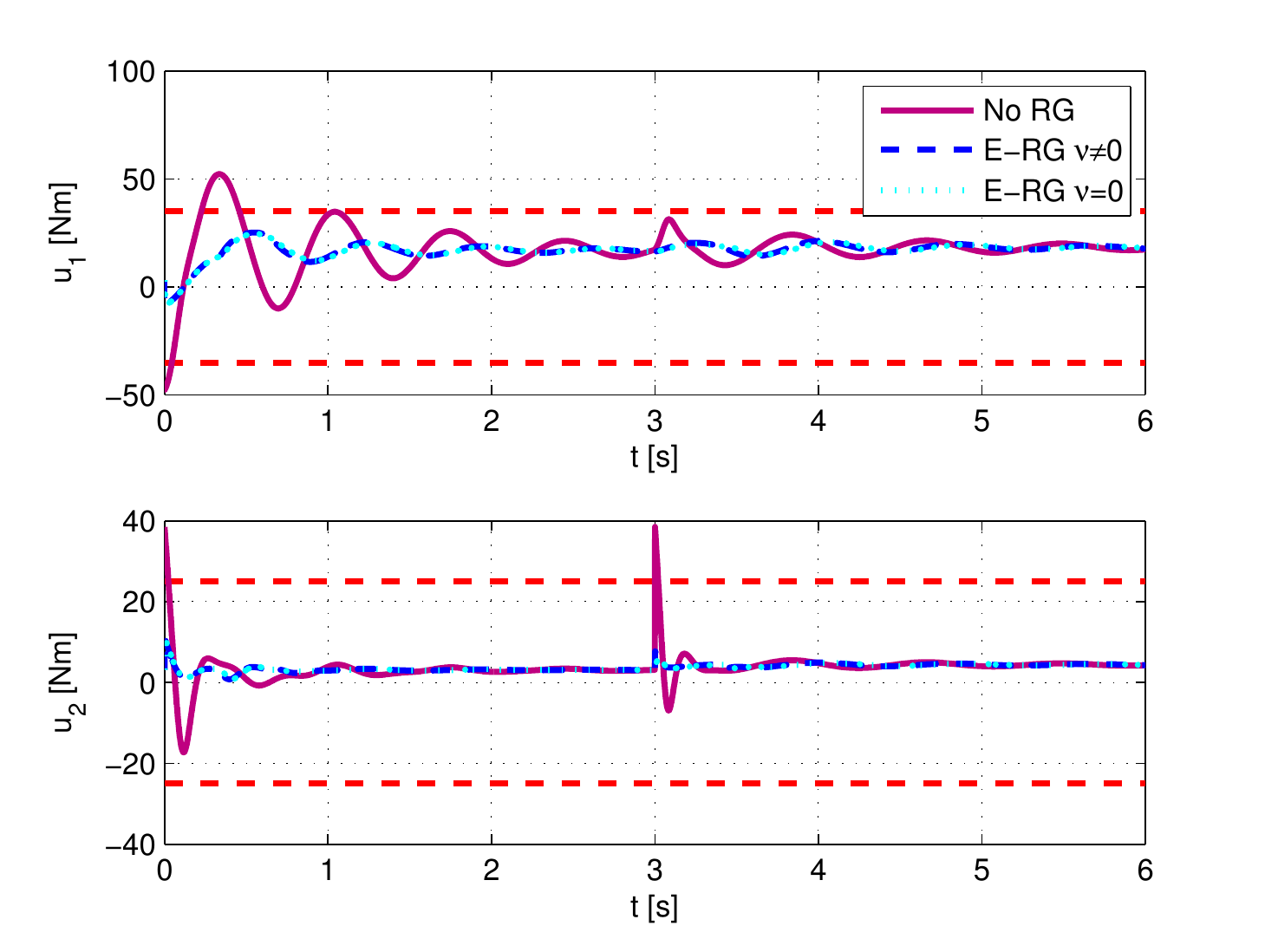}
\vspace{-0.3cm}
{\caption{Applied Input - Robotic Planar Arm} }
\end{figure}
\begin{figure}
\label{fig: gRbt}
\centering
\includegraphics[width=8cm]{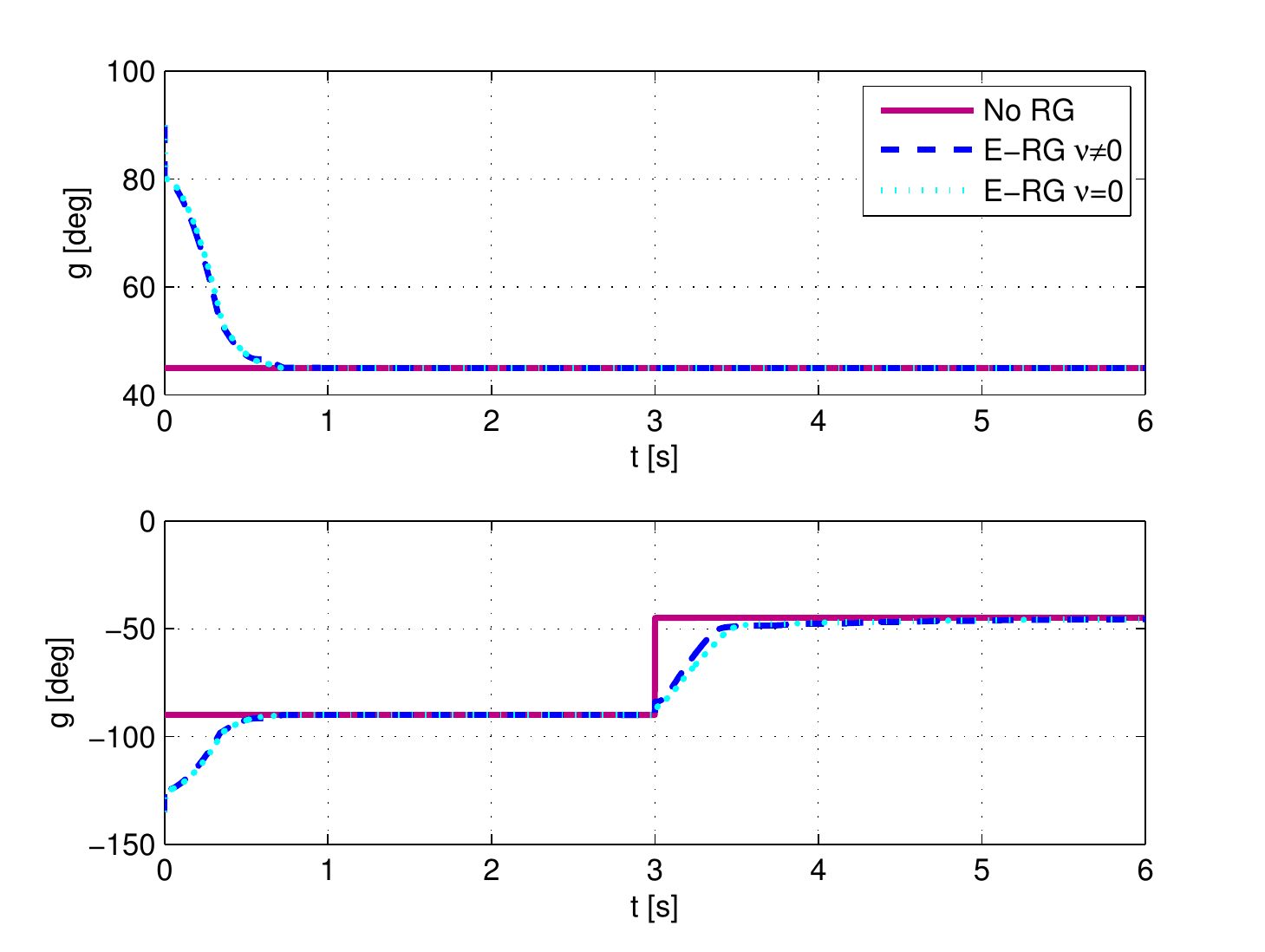}
\vspace{-0.3cm}
{\caption{Applied Reference - Robotic Planar Arm} }
\end{figure}
To implement the E-RG, the mass matrix $M\left(q\right)$ is lower-bounded by
\[
\bar{M}=\left[\begin{array}{cc}
\mu_{1}& \mu_{2}\\
\mu_{2} & \mu_{2}
\end{array}\right] \leq M\left(q\right),
\]
and $\Gamma_i(x_g)$ and $\nu(x,g,r)$ are computed using (\ref{eq: Probot}), (\ref{eq: Gamma}) and (\ref{nurobot}).
Figures 6-8 show the results obtained for
\begin{itemize}
\item \textbf{No RG:} The system is not provided with a Reference Governor and the desired reference is directly applied to the controlled system;
\item \textbf{Explicit RG, with Feedforward:} The E-RG is implemented using  $\epsilon_{1}=10^{-3}, \epsilon_{2}=10^{-3}$ and $\kappa=10^3$. The feedforward term (\ref{eq: nulin}) is used:
\item \textbf{Explicit RG, no Feedforward:} The ERG is implemented with $\nu=0.$
\end{itemize}
As shown in the simulations, the Explicit Reference Governor is successful at enforcing the constraints. Please note that, once again, the difference between the E-RG with and without the feedforward term is marginal.

\section{Experimental Validation}
\begin{figure}
\label{fig: Esp_Setup}
\centering
\includegraphics[width=8cm]{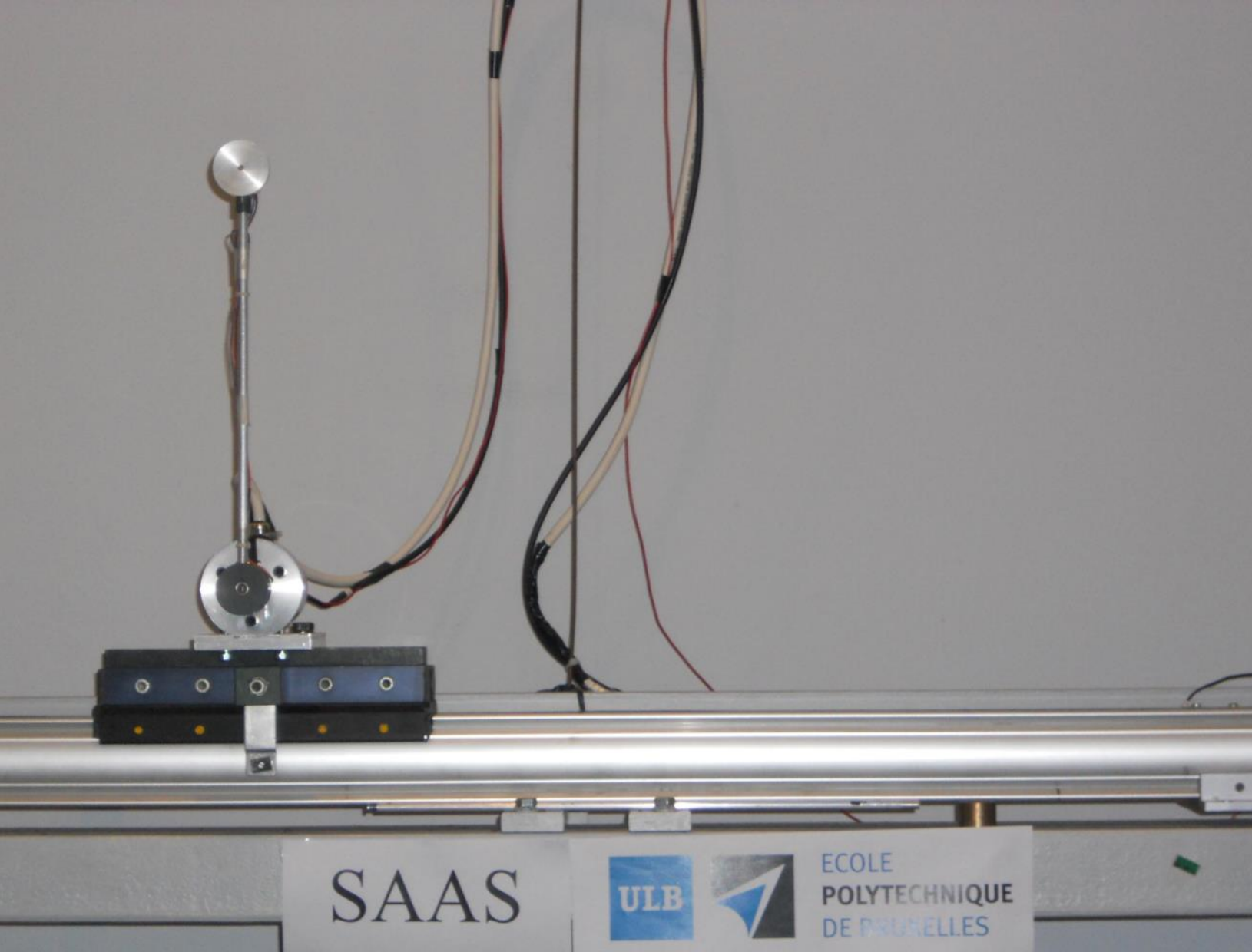}
\vspace{-0.3cm}
{\caption{Experimental set-up of the inverted pendulum on cart} }
\end{figure}
The Explicit Reference Governor has been tested on the inverted pendulum of the SAAS department at ULB. In accordance with the {\em add-on} philosophy of the E-RG, the plant was stabilized using a previously designed Linear Quadratic Regulator (LQR).\\
Under the reasonable assumption that the cart dynamics is not influenced by the pendulum oscillations, the inverted pendulum dynamics is described by the following nonlinear model
\[
\left\{ \begin{array}{l}
\left(M+m\right)\ddot{p}+c_{p}\dot{p}=F\\
\frac{2}{3}l\ddot{\theta}-\ddot{p}\cos\theta-a_g\sin\theta+\frac{2c_{\theta}}{ml}\dot{\theta}=0
\end{array}\right.
\]
where $p$ is the cart position, $\theta$ is the pendulum angle, $F$ is the actuator force, $M,\,m$ are the masses of the cart and of the pendulum, $l$ is the length of the pendulum, $a_g$ is the gravity acceleration and $c_p,\,c_\theta$ are the friction coefficients of the cart rail and of the pendulum hinge. Defining $x=\left[p\,\,\,\dot{p}\,\,\,\theta-\pi\,\,\,\dot{\theta}\right]$, the linearized dynamic model had been identified as
\[
\dot{x}=\left[\begin{array}{cccc}
0 & 1 & 0 & 0\\
0 & -1.92 & 0 & 0\\
0 & 0 & 0 & 1\\
0 & -0.96 & 38.9 & -1.21
\end{array}\right]x+\left[\begin{array}{c}
0\\
22.08\\
0\\
11.04
\end{array}\right]F
\]
which lead to the optimal LQR state feedback $F = [-1.27\,-0.93\,\,\,24.88\,\,\,3.50](x-x_g)$ where $x_g=[1\,\,0\,\,0\,\,0]g$.
The obtained closed-loop system ensures only Regional Asymptotically Stability.\\
The E-RG is therefore tasked with maintaining the state of the system within the basin of attraction as discussed in Remark \ref{RAS}. By using the Lyapunov Function
\[
V=x^{T}\left[\begin{array}{cccc}
6.08 & 3.30 & -66.1 & -6.65\\
3.30 & 2.52 & -53.3 & -5.44\\
-66.1 & -53.3 & 1369 & 124\\
-6.65 & -5.44 & 124 & 14.9
\end{array}\right]x
\]
in conjunction with the nonlinear dynamic model, it is fairly simple to show that $\exists\theta_{\max}:\;|\theta|\leq\theta_{\max}\,\Rightarrow\,\dot{V}<0$.
where $\theta$ is the angular deviation of the pendulum with respect to the upright position. By imposing $\theta_{\max}-\theta\geq0$ and $\theta_{\max}+\theta\geq0$, the Lyapunov constraints $\Gamma_{i}\left(x_g\right),\,i=1,2$ follow from Section \ref{sec: Linear Systems}. This approach, although rigorous, does not take into account parametric uncertainties (e.g. dynamic friction) and unmodeled dynamics (e.g. static friction, actuator inertia) of the real plant. As a result, the Lyapunov constraints were adjusted experimentally. This led to $\Gamma_{i}=210,\,i=1,2$  which implies $\theta_{\max}\approx15^{\circ}$.\\
\begin{figure}
\label{fig: Exp_NoRG}
\centering
\includegraphics[width=8cm]{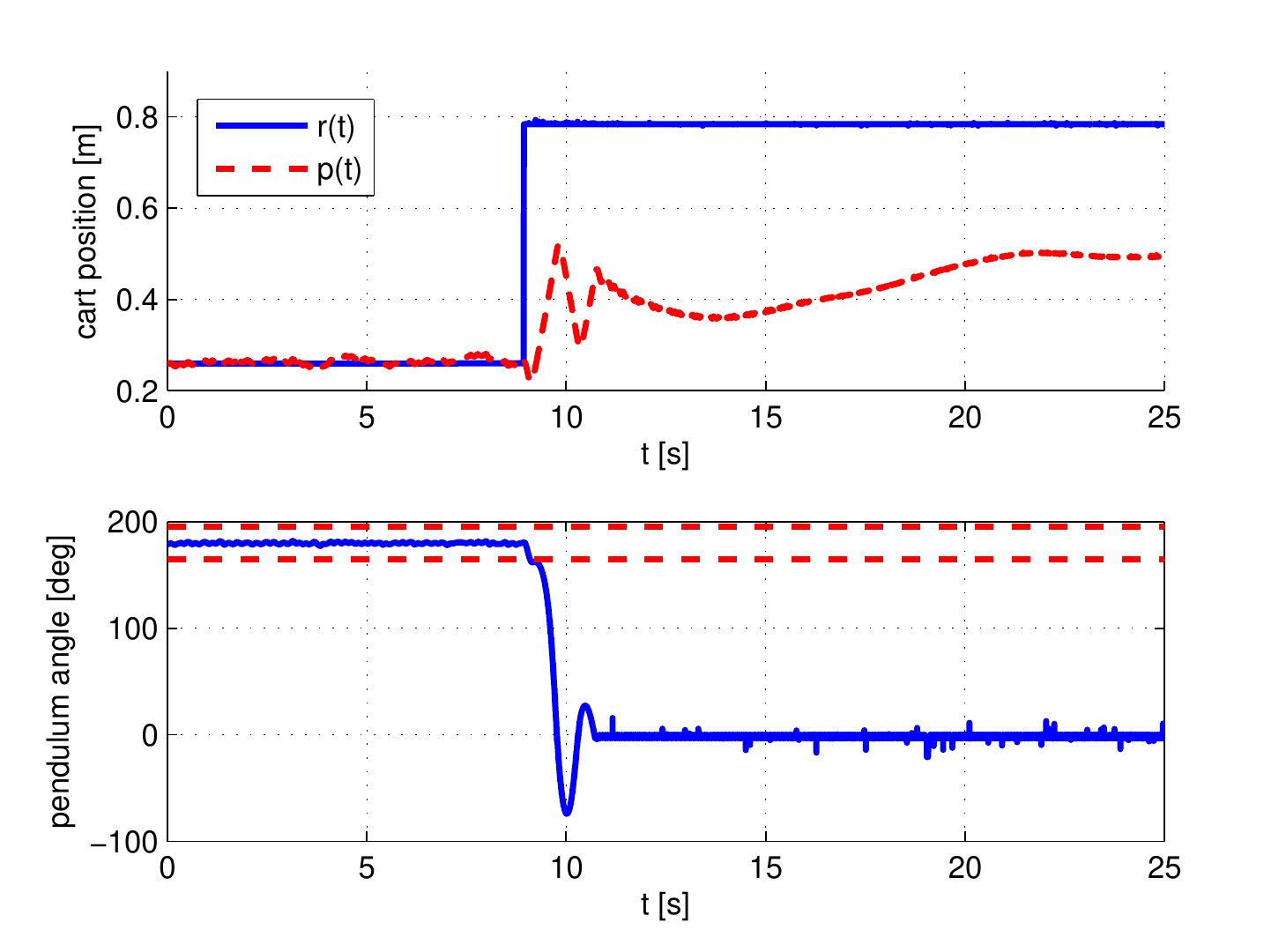}
\vspace{-0.3cm}
{\caption{Response of the closed-loop system to a step variation of the reference} }
\label{fig: Exp_ERG}
\centering
\includegraphics[width=8cm]{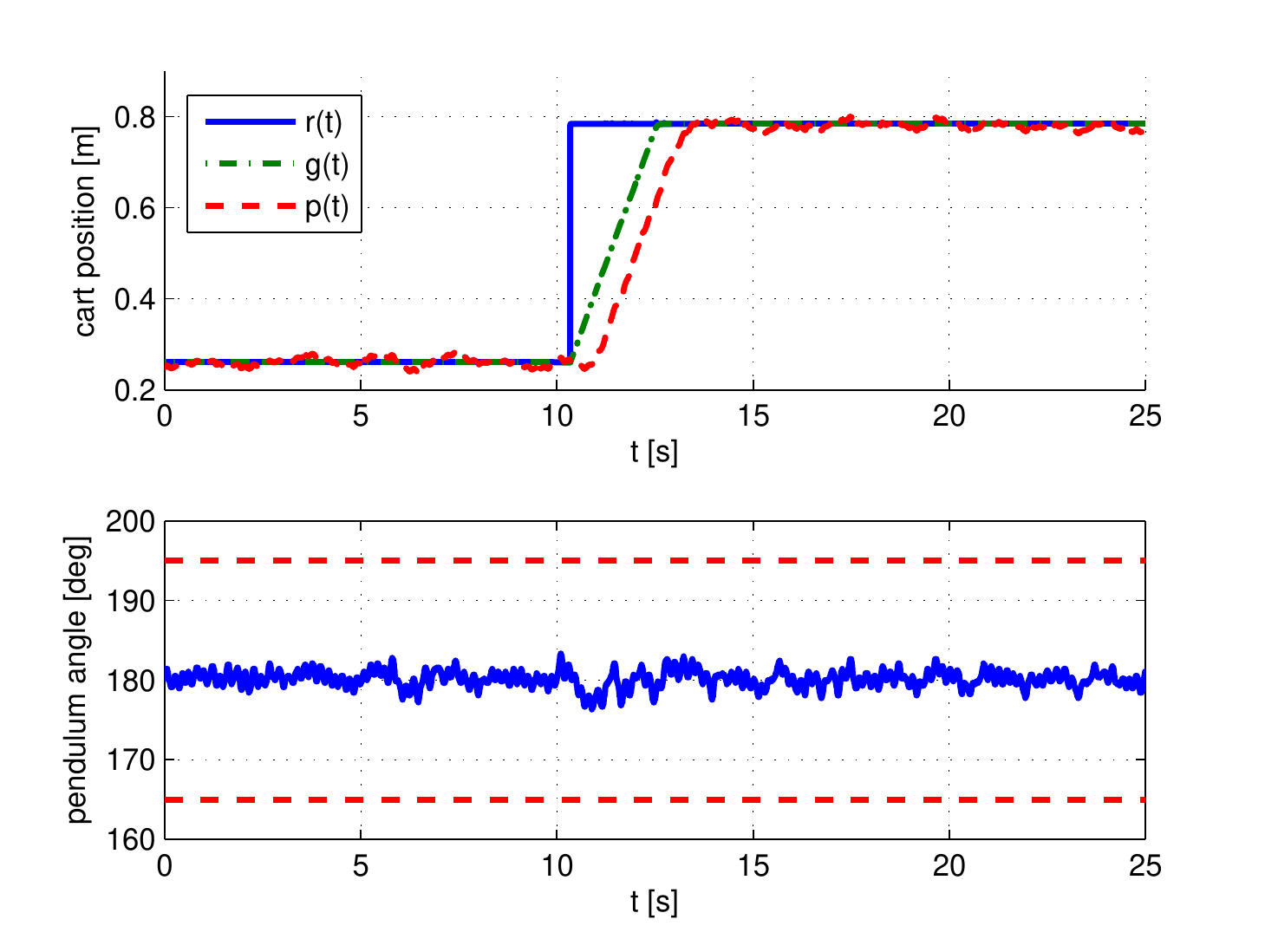}
\vspace{-0.3cm}
{\caption{Response of the closed-loop system integrated with the Explicit Reference Governor} }
\end{figure}
The combined E-RG and LQR were implemented with a sampling rate of $1kHz$. The mean execution time was $0.72ms$, thus confirming that the proposed strategy is computationally inexpensive. The results obtained with and without the E-RG are shown in Figures 9-10. A video of the experiment is available at the URL \url{http://www.gprix.it/papers/InvPen_ERG.wmv}.

\section{Conclusions and Future Works}
In this paper an Explicit Reference Governor is presented which is able to manage the reference of a pre-compensated continuous time system so that a set of convex constraints is not violated. The main ideas of the approach are to convert the state constraints into one constraint on the value of the Lyapunov functions and then to enforce it by suitably acting on the derivative of the applied reference. The feasibility and convergence properties of the scheme are proved. A explicit solution is given for the computation of all the parameters for the relevant case of systems subject to linear constraints and whose Lyapunov function is lower bounded by a quadratic form. Specializations to the cases of linear time-invariant systems and robotic manipulators are also provided.
The authors believe the presented method can be useful in real context as an affordable and cheap way, although not optimal, to manage constraints on nonlinear systems. Future works will focus on extending the presented approach to some classes of nonconvex constraints.

\end{document}